\documentclass[twocolumn]{jpsj2} 
\usepackage{bm}

\title{Fermi-Surface Reconstruction in the Periodic Anderson Model}

\author{Hiroshi \textsc{Watanabe}\thanks{E-mail: hwatanabe@hosi.phys.s.u-tokyo.ac.jp}
and Masao \textsc{Ogata}}

\inst{Department of Physics, University of Tokyo, 7-3-1, Hongo, Bunkyo-ku, Tokyo 113-0033}

\abst{We study ground state properties of periodic Anderson model in a two-dimensional square lattice with variational Monte Carlo method.
It is shown that there are two different types of quantum phase transition: a conventional antiferromagnetic transition   
and a Fermi-surface reconstruction which accompanies a change of topology of the Fermi surface.
The former is induced by a simple back-folding of the Fermi surface while the latter is induced by localization of $f$ electrons.
The mechanism of these transitions and the relation to the recent experiments on Fermi surface are discussed in detail.}

\kword{periodic Anderson model, Fermi-surface reconstruction, heavy-fermion system}

\begin{document}
\maketitle

\section{Introduction} 

Recently, antiferromagnetic (AF) transition which accompanies an abrupt change of Fermi surface has been widely observed in heavy fermion
compounds and has attracted much attention. De Haas-van Alphen (dHvA) experiments in Ce compounds, such as CeRh$_2$Si$_2$~\cite{Araki},
CeIn$_3$~\cite{Settai} and CeRhIn$_5$~\cite{Shishido1},
show that the topology of Fermi surface abruptly changes across the AF transition under pressure. 
According to band calculations, Fermi surface in the AF phase is well explained by a 4$f$-localized model while that in the paramagnetic 
phase is well explained by a 4$f$-itinerant model~\cite{Araki,Settai,Shishido1,Elgazzar}. 
This Fermi-surface change has also been suggested in the Hall effect experiment in YbRh$_2$Si$_2$~\cite{Paschen}.
Although the Hall coefficient shows a continuous change as a function of magnetic field at finite temperature, its slope becomes sharper 
with decreasing temperature. From this, it is claimed that there will be a discontinuous jump of the Hall coefficient at the AF quantum critical point
(QCP) at $T=0$. This means that a change from a ``small Fermi surface'' (AF side) to a ``large Fermi surface'' (paramagnetic side) occurs at QCP. 
In these experiments, the AF transition and the abrupt change of the Fermi surface seem to occur simultaneously.
On the other hand, in CeRh$_{1-x}$Co$_x$In$_5$, an abrupt change of the Fermi surface is observed at $x=0.3\sim0.4$ while the AF QCP exists at 
$x=0.7\sim0.8$~\cite{Ohira-Kawamura,Goh}. In this case, the AF transition and the abrupt change of the Fermi surface are not necessarily coincident 
and the relation between them has not been clarified yet. 

In a conventional picture of the QCP in heavy fermion system, the AF transition is induced by the competition between Kondo screening and 
RKKY interaction~\cite{Doniach}. 
In this case, the conduction electrons ($c$ electrons) and the localized $f$ electrons hybridize with each other and form the composite quasiparticles.
The $f$ electrons obtain the itinerancy through the hybridization and the AF transition is expected to be described by a so-called 
Moriya-Hertz-Millis (MHM) theory~\cite{Moriya,Hertz,Millis}. 
This picture have been extensively studied so far and succeeded in describing the unconventional phenomena such as a 
non-Fermi-liquid behavior around the QCP. However, it is not clear whether the MHM theory can be applied or not to the case where the $f$ electrons 
are strongly localized and the itinerant picture becomes inappropriate. Indeed, the change of the topology of the Fermi surface mentioned above cannot 
be explained by the simple back-folding of the Fermi surface which is based on the itinerant picture.
 
Recently, on the other hand, novel type of theory which considers the localization of $f$ electrons has been proposed by some authors
\cite{Si,Coleman,Senthil,Paul,Pepin}.
It is based on the breakdown of the Kondo screening or the decoupling of $c$ and $f$ electrons. In this case, the change of nature of
the $f$ electrons at the QCP has been emphasized, but the details are still uncertain.

With these facts in mind, we studied the ground state phase diagram of the Kondo lattice model (KLM) in the previous work~\cite{Watanabe}.
This model is one of the simplest models which contain $c$ and $f$ electrons and describe the heavy-fermion system. 
We have found that there are two types of quantum phase transitions: a conventional AF transition and a topological transition of a Fermi surface 
(Fermi-surface reconstruction). The former transition is induced by the folding of the Fermi surface and therefore we denote as ``conventional''.
In the latter transition, on the other hand, the Kondo screening becomes almost irrelevant (but does not disappear) and almost fully polarized AF 
state appears accompanied with the change of the topology of the Fermi surface.

In this paper, we study the ground state properties of the periodic Anderson model (PAM).  
The KLM will not be enough to discuss the issue of the itinerancy and localization of $f$ electrons, since the $f$ electrons are always localized 
in the KLM. 
On the other hand, in the PAM, we can control the number of $f$ electrons and can discuss the change of the nature of $f$ electrons
by changing the parameters such as $U$ (on-site Coulomb interaction), $V$ ($c$-$f$ hybridization) and $E_f$ (energy level of $f$ electrons).
Moreover, we discuss the properties of the Fermi surface in detail by calculating the momentum distribution function which was not discussed in 
the previous paper on the KLM~\cite{Watanabe}.
We show that the Fermi-surface reconstruction observed in the KLM also occurs in the PAM under the condition of $U=\infty$ and sufficiently low 
values of $E_f$.
The Fermi-surface reconstruction have two remarkable aspects as a quantum phase transition. It can be regarded as a kind of Lifshitz transition, 
where the topology of the Fermi surface changes from hole-like to electron-like. At the same time, it can be regarded as a kind of Mott transition, 
where the $f$ electrons change their character from itinerant to localized. 

The organization of this paper is as follows. In \S2, we introduce the model and variational wave functions.
In \S3, we show the results of VMC calculation. The ground state phase diagrams and some physical quantities are studied in detail.
In \S4, we discuss the relation between our results and the experiments.
\S5 is devoted to the summary.

\section{Model and Method}
We study the following PAM in a two-dimensional square lattice,
\begin{multline}
H=\sum_{\bm{k}\sigma}\varepsilon_{\bm{k}}c_{\bm{k}\sigma}^{\dagger}c_{\bm{k}\sigma}
 +E_{f}\sum_{i\sigma}n^f_{i\sigma} \\
 -V\sum_{i\sigma}(f^{\dagger}_{i\sigma}c_{i\sigma}+\mathrm{h.c.})+U\sum_in^f_{i\uparrow}n^f_{i\downarrow}, \label{HM}
\end{multline}
where $\varepsilon_{\bm{k}}$, $E_f$, $V$ and $U$ represent energy dispersion of $c$ electrons, energy level of $f$ electrons, 
hybridization between $c$ and $f$ electrons and on-site Coulomb interaction between $f$ electrons, respectively. 
In the PAM, the number of $f$ electrons at each site is not fixed and therefore the total electron density $n=n_c+n_f$ is a controlling parameter. 

The trial wave function for a paramagnetic state is expressed as,
\begin{equation}
  \left|\Psi \right>  = P^f_{\mathrm{G}} \left|\Phi \right> 
                      = P^f_{\mathrm{G}} \prod_{\bm{k}\sigma}(u_{\bm{k}}c^{\dagger}_{\bm{k}\sigma}
                       +v_{\bm{k}}f^{\dagger}_{\bm{k}\sigma})\left|0\right>
\end{equation}
with
\begin{equation}
 P^f_{\mathrm{G}}= \prod\left(1-(1-g)n^f_{i\uparrow}n^f_{i\downarrow}\right), 
\end{equation}
and
\begin{align}
 u^2_{\bm{k}}&=\frac{1}{2}\left[1-\frac{\varepsilon_{\bm{k}}-\tilde{E_f}}{\sqrt{(\varepsilon_{\bm{k}}-\tilde{E_f})^2+4\tilde{V}^2}}\right], \label{u} \\
 v^2_{\bm{k}}&=\frac{1}{2}\left[1+\frac{\varepsilon_{\bm{k}}-\tilde{E_f}}{\sqrt{(\varepsilon_{\bm{k}}-\tilde{E_f})^2+4\tilde{V}^2}}\right]. \label{v}
\end{align}
Here $u_{\bm{k}}$ $(v_{\bm{k}})$ corresponds to the mixing amplitude of $c$ ($f$) electrons. 
$P^f_{\mathrm{G}}$ is a Gutzwiller projection operator which reduces the probability of double occupancy of $f$ electrons at the same site.
Since we consider the case of $U=\infty$, we set $g=0$ in the following.
The unprojected wave function $\left|\Phi \right>$ is obtained by diagonalizing the following one-body Hamiltonian,
\begin{equation}
 H=\sum_{\bm{k}\sigma}
    \left(c_{\bm{k}\sigma}^{\dagger}, f_{\bm{k}\sigma}^{\dagger}\right)
  \begin{pmatrix}
   \varepsilon_{\bm{k}} & -\tilde{V} \\ -\tilde{V} & \tilde{E_f}
 \end{pmatrix}
 \begin{pmatrix}
   c_{\bm{k}\sigma} \\ f_{\bm{k}\sigma}
 \end{pmatrix} \label{PM}.
\end{equation}
It is just the $U=0$ PAM with ``effective'' hybridization $\tilde{V}$ and ``effective'' $f$-electron level $\tilde{E_f}$. 
$\tilde{V}$ and $\tilde{E_f}$ are variational parameters and they are optimized so as to minimize the variational energy.
This form of trial wave function is also used in ref.~\citen{Shiba1} for a one-dimensional case. 
As an AF trial wave function, we construct a state $\left|\Phi \right>$ by diagonalizing the following Hamiltonian, 
\begin{align}
H&=\sum_{\bm{k}\sigma}
        \left( c_{\bm{k}\sigma}^{\dagger}, c_{\bm{k}+\bm{Q}\sigma}^{\dagger},  
   f_{\bm{k}\sigma}^{\dagger}, f_{\bm{k}+\bm{Q}\sigma}^{\dagger}\right)  \times \notag \\
  &  \begin{pmatrix}
   \varepsilon_{\bm{k}} & \sigma \Delta_c & -\tilde{V} & 0 \\ 
   \sigma \Delta_c & \varepsilon_{\bm{k}+\bm{Q}} & 0 & -\tilde{V} \\
   -\tilde{V} & 0 & \tilde{E_f} & -\sigma \Delta_f \\
   0 & -\tilde{V} & -\sigma \Delta_f & \tilde{E_f} 
 \end{pmatrix}
 \begin{pmatrix}
  c_{\bm{k}\sigma} \\ c_{\bm{k}+\bm{Q}\sigma} \\ 
  f_{\bm{k}\sigma} \\ f_{\bm{k}+\bm{Q}\sigma}    
 \end{pmatrix}, \label{AF}
\end{align} 
with $\bm{Q}=(\pi,\pi)$ being an AF ordering vector.
Here $\Delta_c$ and $\Delta_f$ represent the variational parameters of AF gap in $c$ and $f$ electrons, respectively.
The summation over $\bm{k}$ is carried out in the folded AF Brillouin zone.
If we set $\Delta_c=\Delta_f=0$, eq.~(\ref{AF}) reduces to eq.~(\ref{PM}).
Although the choice of the basis and the notation of variational parameters are slightly different from those used in ref.~\citen{Watanabe}, 
the obtained unprojected band dispersion is the same with the previous paper~\cite{Watanabe}
 ($\Delta_f$ and $\Delta_c$ corresponds to $M$ and $m$ in ref.~\citen{Watanabe}).

We optimize the four variational parameters $(\tilde{V},\tilde{E_f},\Delta_f,\Delta_c)$ and find the lowest energy state 
in a $V$-$n$ phase diagram. 
Possible candidates for the ground state are classified according to the shape of the Fermi surface and the band dispersion of the 
one-body part $\left|\Phi \right>$: 
paramagnetic metal (PM), AF metal with hole-like Fermi surface (AF$_{\mathrm{h}}$), AF metal with electron-like Fermi surface
(AF$_{\mathrm{e}}$) and AF metal with $\tilde{V}=0$ (AF$_{\mathrm{S}}$, S denotes ``small").
This notation is the same with that in the previous paper~\cite{Watanabe} and the schematic Fermi surfaces are shown in Fig.~\ref{fig1}. 
In PM, AF$_{\mathrm{h}}$ and AF$_{\mathrm{e}}$, $\tilde{V}$ is finite and the one-body part forms the $c$-$f$ hybridized band.
It is based on the concept of ``large'' Fermi surface. In contrast, $\tilde{V}=0$ in AF$_{\mathrm{S}}$, and $c$ and $f$ electrons form their
band dispersions individually. AF$_{\mathrm{S}}$ is based on the concept of ``small'' Fermi surface.
However, note that the expressions of ``large'' and ``small'' Fermi surface are not appropriate in the AF state 
since the volumes of the Fermi surface are the same in AF$_{\mathrm{h}}$, AF$_{\mathrm{e}}$ and AF$_{\mathrm{S}}$.
Moreover, AF$_{\mathrm{e}}$ and AF$_{\mathrm{S}}$ have the same Fermi-surface topology and cannot be distinguished from each other only by the 
Fermi surface. The difference between them are discussed in the next section.

\section{Results}
\subsection{Ground State Phase Diagram}
\begin{figure*}
\begin{center}
\includegraphics[width=17cm]{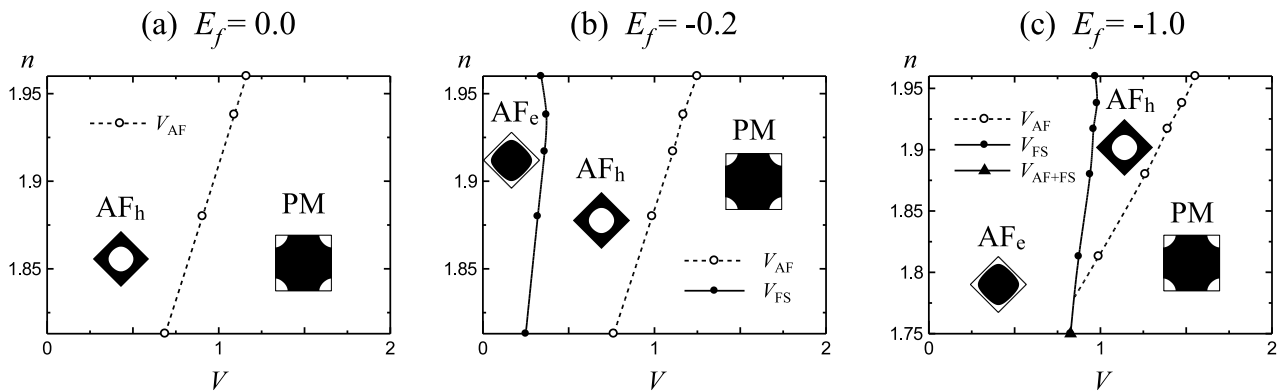}
\caption{Ground state phase diagram of $U=\infty$ PAM in a two-dimensional square lattice. Solid curves ($V_{\mathrm{FS}}$ and
$V_{\mathrm{AF+FS}}$) represent the first-order phase transition and dotted curves ($V_{\mathrm{AF}}$) represent the second-order one.}
\label{fig1}
\end{center}
\end{figure*}

We show the ground state phase diagrams for different values of $E_f$ in Fig.~\ref{fig1}.
The system sizes are set to be 8$\times$8, 10$\times$10 and 12$\times$12 with periodic-antiperiodic boundary condition. 
When $E_f=0.0$, the state changes from PM to AF$_{\mathrm{h}}$ at $V=V_{\mathrm{AF}}$ as shown in Fig.~\ref{fig1}(a).
The first Brillouin zone is folded and the AF gap opens along the newly formed AF Brillouin zone boundary.
This is a ``conventional'' second-order AF transition.
The region of AF$_{\mathrm{h}}$ extends to $V\rightarrow0$.
On the other hand, when $E_f=-0.2$, a novel type of quantum phase transition occurs and AF$_{\mathrm{e}}$ is realized
for $V<V_{\mathrm{FS}}$, as shown in Fig~\ref{fig1}(b).
At $V=V_{\mathrm{FS}}$, the band dispersion discontinuously changes from convex upward (AF$_{\mathrm{h}}$ side) to convex downward
(AF$_{\mathrm{e}}$ side).
It is regarded as a kind of Lifshitz transition, where the topology of the Fermi surface changes.
We call this transition a ``Fermi-surface reconstruction'' and its origin is discussed in \S3.2 in detail.
As $E_f$ becomes lower, the region of AF$_{\mathrm{e}}$ becomes larger and finally overlaps the region of AF$_{\mathrm{h}}$.
As a result, when $E_f=-1.0$ (Fig.~\ref{fig1}(c)), the direct transition from PM to AF$_{\mathrm{e}}$ appears for $n\lesssim1.78$.
We can see that the stability of AF$_{\mathrm{e}}$ greatly depends on the value of $E_f$.

\begin{figure*}
\begin{center}
\includegraphics[width=17cm]{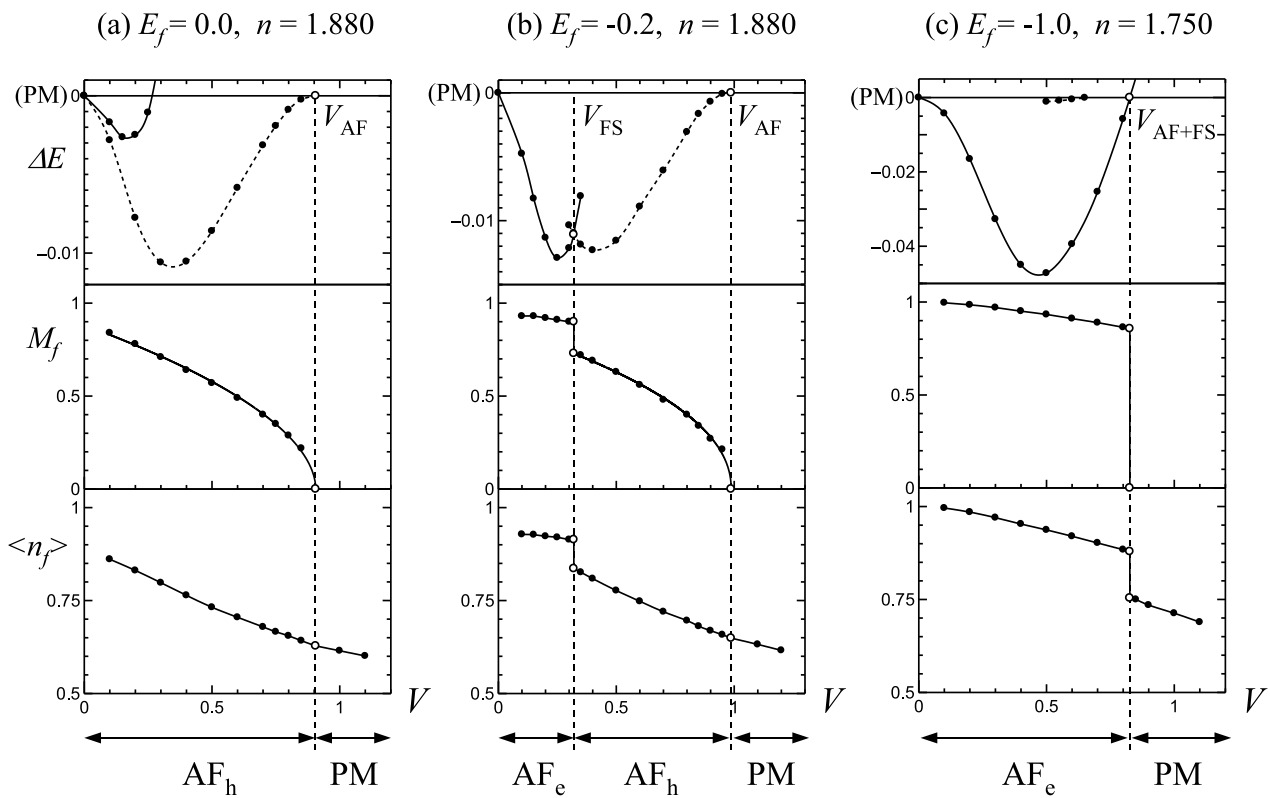}
\caption{$V$ dependence of condensation energy $\Delta E$, staggered magnetization of $f$ electrons $M_f$ and average $f$-electron density
$\left<n_f\right>$. $\Delta E$ are variational energies of AF$_{\mathrm{e}}$ (solid curves) and AF$_{\mathrm{h}}$ (dotted curves) compared 
with that of PM. Open circles correspond to the transition points.}
\label{fig2}
\end{center}
\end{figure*}

For fixed values of $E_f$ and $n$, the change of the ground state is classified into three cases:
(i) PM $\rightarrow$ AF$_{\mathrm{h}}$, (ii) PM $\rightarrow$ AF$_{\mathrm{h}}$ $\rightarrow$ AF$_{\mathrm{e}}$ and 
(iii) PM $\rightarrow$ AF$_{\mathrm{e}}$. 
In Fig.~\ref{fig2}, we show the energy comparison between the trial wave functions for each case and discuss the behaviors of the staggered 
magnetization of 
$f$ electrons, $M_f=\bigl<n^f_{\mathrm{A\uparrow}}\bigr>-\bigl<n^f_{\mathrm{A\downarrow}}\bigr>
                                     =-\bigl(\bigl<n^f_{\mathrm{B\uparrow}}\bigr>-\bigl<n^f_{\mathrm{B\downarrow}}\bigr>\bigr)$,
and the average $f$-electron density $\left<n_f\right>$.

Figure~\ref{fig2}(a) shows the case of (i) at $E_f=0.0$ and $n=1.880$. 
The ground state is PM for $V>V_{\mathrm{AF}}$ and AF$_{\mathrm{h}}$ for $V<V_{\mathrm{AF}}$. 
The variational energy of AF$_{\mathrm{h}}$ smoothly merges with that of PM (horizontal axis) at $V=V_{\mathrm{AF}}$, 
indicating the second-order (continuous) AF transition. 
$M_f$ continuously increases from zero as $V$ decreases from $V_{\mathrm{AF}}$. 
$\left<n_f\right>$ monotonically increases as $V$ decreases from the PM side, and shows continuous change across the AF transition.

The case of (ii) at $E_f=-0.2$ and $n=1.880$ is shown in Fig.~\ref{fig2}(b). 
As in the case of (i), the transition from PM to AF$_{\mathrm{h}}$ at $V=V_{\mathrm{AF}}$ is second-order. 
The changes of $M_f$ and $\left<n_f\right>$ across the AF transition are continuous.
On the other hand, when the Fermi-surface reconstruction (AF$_{\mathrm{h}}$ $\rightarrow$ AF$_{\mathrm{e}}$) occurs, abrupt changes are observed. 
At $V=V_{\mathrm{FS}}$, the energy crossing occurs and the ground state changes from AF$_{\mathrm{h}}$ to AF$_{\mathrm{e}}$.
It is clearly a first-order transition and physical quantities such as $M_f$ and $\left<n_f\right>$ discontinuously increase across the transition.
As one can see from Fig.~\ref{fig2}(b), the $f$ electrons are almost localized in AF$_{\mathrm{e}}$ and $M_f\simeq\left<n_f\right>$ is satisfied.
It shows that the Kondo screening, which works so as to suppress the AF order, becomes almost irrelevant in AF$_{\mathrm{e}}$ and
the local $f$-electron moment is almost fully polarized.

The case of (iii) at $E_f=-1.0$ and $n=1.750$ is shown in Fig.~\ref{fig2}(c). Contrary to the cases of (i) and (ii), AF$_{\mathrm{h}}$ does not appear
for any values of $V$; the variational energy of AF$_{\mathrm{h}}$ is always higher than that of AF$_{\mathrm{e}}$, and
the ground state directly changes from PM to AF$_{\mathrm{e}}$ at $V=V_{\mathrm{AF+FS}}$. It is a first-order transition, where the AF transition 
and the Fermi-surface reconstruction occur simultaneously.
As a result, $M_f$ shows discontinuous jump from zero to a finite value at $V=V_{\mathrm{AF+FS}}$.
As in the case of (ii), $M_f\simeq\left<n_f\right>$ is satisfied and the almost fully polarized AF state is realized in AF$_{\mathrm{e}}$.

Note that AF$_{\mathrm{S}}$ does not appear for all cases. It is a natural result since AF$_{\mathrm{S}}$ cannot gain the mixing energy because of
$\tilde{V}=0$ and it is energetically unfavorable under the condition of $V\neq0$.
The introduction of small but finite value of $\tilde{V}$ does
not change the shape of the Fermi surface very much, but it leads to a mixing energy gain and stabilizes AF$_{\mathrm{e}}$.
In AF$_{\mathrm{e}}$, the $c$ electrons enlarge the occupied phase space and become heavy compared with the case of AF$_{\mathrm{S}}$ 
because of the finite value of $\tilde{V}$. 
Namely, AF$_{\mathrm{e}}$ and AF$_{\mathrm{S}}$ have almost the same Fermi surface but their internal structures of quasiparticles
are completely different.

\subsection{Origin of the Fermi Surface Reconstruction}
\begin{figure}[t!]
\begin{center}
\includegraphics[width=7.5cm]{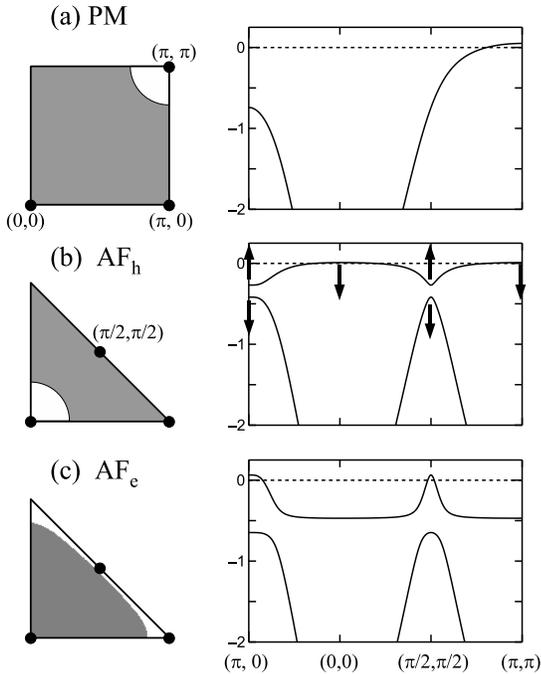}
\caption{Band dispersions of PM ($V$=1.50), AF$_{\mathrm{h}}$ ($V$=1.00) and AF$_{\mathrm{e}}$ ($V$=0.70) at $E_F=-1.0$ and $n=1.880$. 
The Fermi energy is set to be 0.
Arrows indicate the direction of the change of dispersion when $V$ decreases (see text).}
\label{fig3}
\end{center}
\end{figure}

In \S3.1, we have shown the existence of the novel type of quantum phase transition of ``Fermi-surface reconstruction''.
In this section, we discuss its origin.
Figure~\ref{fig3} shows the band dispersions of the three different states obtained by diagonalizing the one-body part 
$\left|\Phi \right>$ with optimized parameters: 
PM ($V$=1.50), AF$_{\mathrm{h}}$ ($V$=1.00) and AF$_{\mathrm{e}}$ ($V$=0.70) at $E_F=-1.0$ and $n=1.880$.
In PM, only the lower hybridized band is shown since the upper one is away from the Fermi energy and unoccupied.
In AF$_{\mathrm{h}}$, the energy band is folded and the AF gap opens along the AF Brillouin zone boundary 
(($\pi,0$) and ($\pi/2,\pi/2$) in Fig.~\ref{fig3}(b)).
This folding does not change the topology of the Fermi surface compared with Fig.~\ref{fig3}(a).
As $V$ decreases in AF$_{\mathrm{h}}$, the optimized value of $\tilde{V}$ also decreases and that of $\Delta_f$ increases. 
Following the growth of the AF gap, the band dispersion changes in the direction as shown by the arrows in Fig.~\ref{fig3}(b) 
and finally changes into a dispersion which is convex downward. 
As a result, quasiparticle occupation changes in the $\bm{k}$-space and the Fermi surface is reconstructed from hole-like (AF$_{\mathrm{h}}$)
to electron-like (AF$_{\mathrm{e}}$) at $V=V_{\mathrm{FS}}$.
According to this mechanism, it is in principle possible that the Fermi-surface reconstruction becomes a second-order transition. 
However, within our trial wave function, the Fermi-surface reconstruction is clearly a first-order transition which accompanies the energy crossing.

Next we discuss the origin of the Fermi-surface reconstruction from a viewpoint of energy gain.
The variational parameter $\tilde{V}$ corresponds to the effective hybridization and thus it will be directly related to the 
degree of Kondo screening. It works so as to gain the mixing energy $E_{\mathrm{mix}}$, the third term in eq.(\ref{HM}).
On the other hand, $\Delta_f$ corresponds to the AF gap and controls the magnitude of the AF order. 
It stabilizes the local $f$-electron moment and suppresses the hybridization, 
leading to gain the kinetic energy of $c$ electrons $E_{\mathrm{kin}}$ and the $f$-electrons energy 
$E_{n_f}$, which are the first and the second terms in eq.(\ref{HM}), respectively.
In PM and AF$_{\mathrm{h}}$, the role of $\tilde{V}$ is dominant and the system forms the hole-like Fermi surface which is favorable 
for $E_{\mathrm{mix}}$. 
In AF$_{\mathrm{e}}$, on the other hand, the role of $\Delta_f$ exceeds that of $\tilde{V}$ and the electron-like Fermi surface, which is
favorable for $E_{\mathrm{kin}}$ and $E_{n_f}$, is formed. 
Namely, the origin of the Fermi-surface reconstruction is a change of the mechanism of the energy gain.
In AF$_{\mathrm{e}}$, the localization of $f$ electrons is essential for the energy gain and the AF order is considered to be additional.
Therefore, the region of AF$_{\mathrm{e}}$ in the phase diagram mainly depends on $E_f$, but not so much on the total electron density $n$  
which is closely related to the stability of the AF order.
It is in contrast to the case of AF$_{\mathrm{h}}$ whose stability greatly depends on $n$.
 
The existence of the Fermi-surface reconstruction is also suggested in the recent study of the cellular dynamical mean-field theory (CDMFT) by
De Leo \textit{et al}~\cite{Leo}. 
They have calculated spectral functions for paramagnetic and AF states and shown that the AF transition is accompanied by a dramatic rearrangement
of the spectral weight. 
It suggests the change of the topology of the Fermi surface and can be regarded as a kind of Fermi-surface reconstruction. 
Although the condition of their calculation is somewhat different with ours (the chemical potential is fixed in ref.~\citen{Leo} instead of the 
electron density), their result is consistent with the existence of the Fermi-surface reconstruction. 

Let us mention here the order of the transition. In our results, the Fermi-surface reconstruction is clearly a first-order transition.
However, as mentioned before, it is to be noted that the second-order (or other continuous) transition cannot be completely ruled out 
since the result depends on the choice of the trial wave functions.

Finally, we compare the present results with the Kondo lattice model (KLM).
In the previous paper~\cite{Watanabe}, we have shown that the Fermi-surface reconstruction occurs in the KLM.
The result that the Fermi-surface reconstruction occurs for $U=\infty$ and sufficiently low value of $E_f$ in the PAM is consistent with 
the result of the KLM, 
since the KLM is an effective model of the PAM in the limit of $U\rightarrow\infty$ and $n_f\rightarrow 1$.
Recent studies of the dynamical cluster approximation~\cite{Martin} and the Gutzwiller approximation~\cite{Lanata} also support the existence of the 
Fermi-surface reconstruction in
the KLM. In the Gutzwiller approximation, the one-body part is more improved than ours but the obtained phase diagram is in good agreement with ours.
This means that our variational wave functions capture the essence of the AF transition and the Fermi-surface reconstruction in spite of their rather
simple forms.
 
\subsection{Momentum Distribution Function}
So far, we have discussed the Fermi surface obtained from the one-body part $\left|\Phi \right>$ without projection. However, it is in principle
possible that the Fermi surface changes by the effect of the projection operator, $P^f_{\mathrm{G}}$.
Since the Fermi surface is determined from the discontinuities in the momentum distribution function,
we calculate them for $c$ and $f$ electrons in this section.
They are defined as,
\begin{align}
 n_c(\bm{k})&=\frac{1}{2}\sum_{\sigma}\big<c^{\dagger}_{\bm{k}\sigma}c_{\bm{k}\sigma}\big> 
             =\frac{1}{2N}\sum_{i,j,\sigma}\mathrm{e}^{\mathrm{i}\bm{k}\cdot(\bm{r}_i-\bm{r}_j)}
              \big<c^{\dagger}_{i\sigma}c_{j\sigma}\big> \\
 n_f(\bm{k})&=\frac{1}{2}\sum_{\sigma}\big<f^{\dagger}_{\bm{k}\sigma}f_{\bm{k}\sigma}\big> 
             =\frac{1}{2N}\sum_{i,j,\sigma}\mathrm{e}^{\mathrm{i}\bm{k}\cdot(\bm{r}_i-\bm{r}_j)}
              \big<f^{\dagger}_{i\sigma}f_{j\sigma}\big>.
\end{align}
We calculate $n_c(\bm{k})$ and $n_f(\bm{k})$ with and without projection, $P^f_G$, in order to study the role of the projection operator.
We set $E_f=-1.0$ and $n=n_c+n_f=612/324=1.889$. Each $\bm{k}$ point is
shifted along $y$ direction by $\delta=\pi/9$ because of the antiperiodic boundary condition.

\begin{figure}[t!]
\begin{center}
\includegraphics[width=7.8cm]{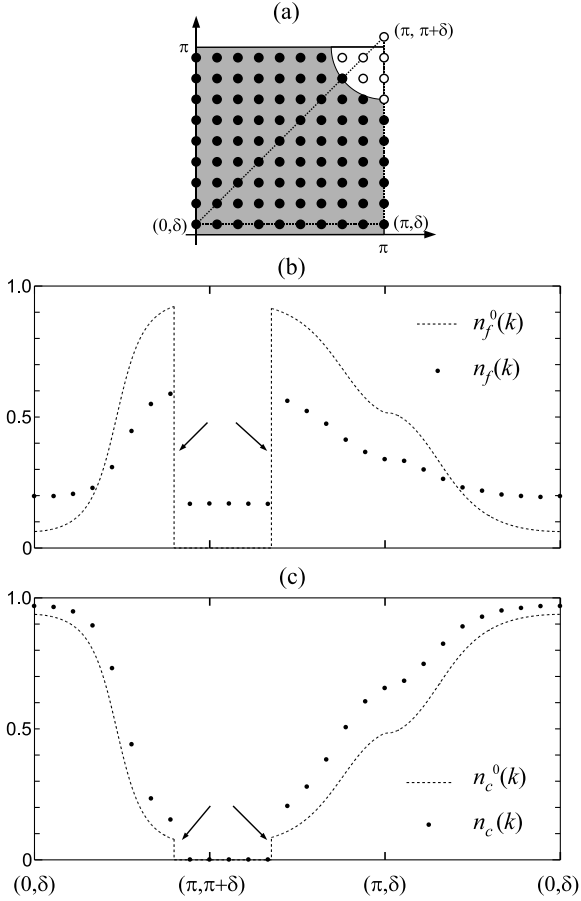}
\caption{Momentum distribution functions of PM at $E_F=-1.0$, $n=1.889$ and $V=1.50$. (a) Occupation of quasiparticles and Fermi surface in the 
first Brillouin zone. Solid (open) circles represent the occupied (unoccupied) $\bm{k}$-points. 
Each $\bm{k}$ point is shifted along $y$ direction by $\delta=\pi/9$ because of the antiperiodic boundary condition.
Calculated momentum distribution functions
of (b) $f$ electrons and (c) $c$ electrons with and without ($^0$) projection. Arrows indicate the positions of discontinuities in 
$n_f^0(\bm{k})$ and $n_c^0(\bm{k})$. The optimized variational parameters are 
$(\tilde{V},\tilde{E_f},\Delta_f,\Delta_c)=(1.0899,-0.0460,0.0000,0.0000)$} 
\label{fig4}
\end{center}
\end{figure}

Figure~\ref{fig4} shows the result for PM at $V=1.50$. 
In PM, the momentum distribution function of $c$ and $f$ electrons without projection are identical to $u_{\bm{k}}^2$ (eq.(\ref{u})) and 
$v_{\bm{k}}^2$ (eq.(\ref{v})). 
We denote them as $n_c^0(\bm{k})$ and $n_f^0(\bm{k})$, respectively.
As shown by arrows in Fig.~\ref{fig4}(b) and (c), 
$n_c^0(\bm{k})$ and $n_f^0(\bm{k})$ have discontinuities at the same position.
Figure~\ref{fig4}(a) shows the Fermi surface determined from $n_c^0(\bm{k})$ and $n_f^0(\bm{k})$.
We call it ``original Fermi surface'' in the following.

As shown in Fig.~\ref{fig4}(b), the projection operator flattens $n_f(\bm{k})$ in the whole Brillouin zone and the discontinuities are fairly reduced. 
Moreover, $n_f(\bm{k})$ has finite values even outside the original Fermi surface. 
They are known as a typical role of the on-site Coulomb interaction $U$. 
Judging from the $\bm{k}$-dependence of $n_f(\bm{k})$ near the original Fermi surface, it is probable that the position of the 
discontinuities does not change in the presence of $P^f_{\mathrm{G}}$.
Contrary to $n_f(\bm{k})$, $n_c(\bm{k})$ is generally enhanced by projection since the excluded $f$ electrons come into the $c$-electron orbital. 
As shown in Fig.~\ref{fig4}(c), the discontinuities in $n_c(\bm{k})$ are rather uncertain, but when we fit the data points 
by polynomials, we find that there are finite jumps at the original Fermi surface. 
Although it cannot be completely excluded that the discontinuities in $n_c(\bm{k})$ disappear, we expect that the original Fermi surface is 
preserved.
If this is the case, we can say that our trial wave function satisfies the Luttinger's theorem: the volume of the Fermi surface is 
determined by the electron density and not affected by the strength of electron correlation, unless the phase transition occurs. 

Next we discuss the AF state. As in the case of PM, $n_c^0(\bm{k})$ and $n_f^0(\bm{k})$ are explicitly calculated from the 
one-body part $\left|\Phi \right>$. By diagonalizing the $4\times4$ matrix in eq.(\ref{AF}), we obtain four quasiparticles 
($\alpha_{\bm{k}\sigma},\beta_{\bm{k}\sigma},\gamma_{\bm{k}\sigma},\delta_{\bm{k}\sigma}$) and band dispersions 
($E^{\alpha}_{\bm{k}},E^{\beta}_{\bm{k}},E^{\gamma}_{\bm{k}},E^{\delta}_{\bm{k}}$) in the folded AF Brillouin zone.
In the present condition, the Fermi energy crosses only the $\beta$-band (see Figs.~\ref{fig3}(b) and (c)).  
The $\alpha$-band is completely filled while the $\gamma$- and the 
$\delta$-band are empty, i.e., $n^{\alpha}_{\bm{k}}=1$ and $n^{\gamma}_{\bm{k}}=n^{\delta}_{\bm{k}}=0$ for all values of $\bm{k}$. 
Then $n_c^0(\bm{k})$ and $n_f^0(\bm{k})$ are expressed as
\begin{equation}
 \begin{cases}
  \;n_c^0(\bm{k})&=       A_1(\bm{k})+\theta(\varepsilon_{\mathrm{F}}-E^{\beta}_{\bm{k}})A_5(\bm{k}) \\
  \;n_c^0(\bm{k}+\bm{Q})&=A_2(\bm{k})+\theta(\varepsilon_{\mathrm{F}}-E^{\beta}_{\bm{k}})A_6(\bm{k}) \\
  \;n_f^0(\bm{k})&=       A_3(\bm{k})+\theta(\varepsilon_{\mathrm{F}}-E^{\beta}_{\bm{k}})A_7(\bm{k}) \\
  \;n_f^0(\bm{k}+\bm{Q})&=A_4(\bm{k})+\theta(\varepsilon_{\mathrm{F}}-E^{\beta}_{\bm{k}})A_8(\bm{k})
 \end{cases},
 \label{nk0-AF}
\end{equation}
where $\theta(x)$ is the step function and $\varepsilon_{\mathrm{F}}$ denotes the Fermi energy.
Here $A_i(\bm{k})$ are derived from the eigenvectors obtained in the diagonalization.
We can see that the discontinuities in $n_c^0(\bm{k})$ and $n_f^0(\bm{k})$ are determined by the occupation of the $\beta$-band.  

\begin{figure}[t!]
\begin{center}
\includegraphics[width=7.8cm]{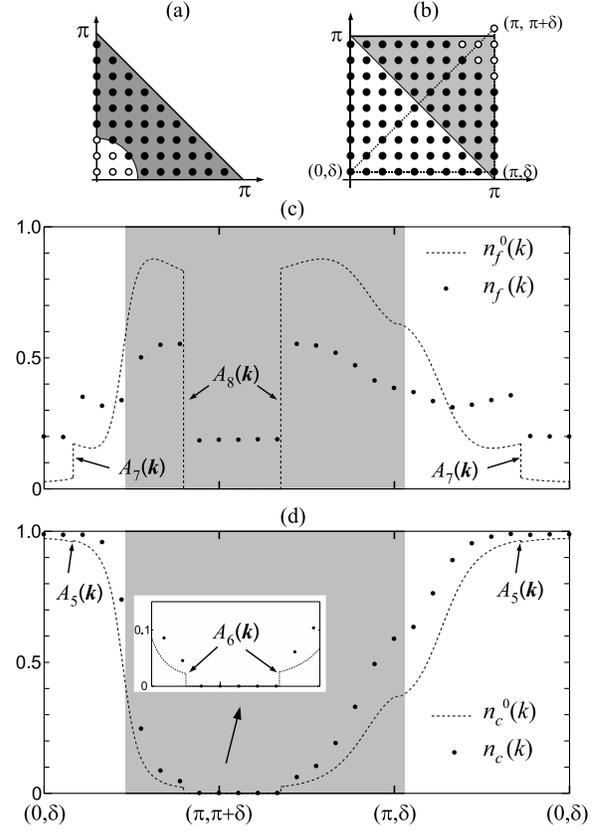}
\caption{Momentum distribution functions of AF$_{\mathrm{h}}$ at $E_F=-1.0$, $n=1.889$ and $V=1.00$. (a) Occupation of quasiparticles in the $\beta$-band
and Fermi surface in the folded AF Brillouin zone. Solid (open) circles represent the occupied (unoccupied) $\bm{k}$-points. 
(b) First Brillouin zone described with the extended zone scheme.
Calculated momentum distribution functions of (c) $f$ electrons and (d) $c$ electrons with and without ($^0$) projection. 
Arrows indicate the positions of discontinuities in $n_f^0(\bm{k})$ and $n_c^0(\bm{k})$. The optimized variational parameters are 
$(\tilde{V},\tilde{E_f},\Delta_f,\Delta_c)=(0.6368,-0.3236,0.1193,0.0000)$} 
\label{fig5}
\end{center}
\end{figure}

Figure~\ref{fig5} shows the result for AF$_{\mathrm{h}}$ at $V=1.00$.
The original Fermi surface determined from $n_c^0(\bm{k})$ and $n_f^0(\bm{k})$ is shown in Fig.~\ref{fig5}(a).
To make the comparison with the case of PM easier, we use the extended Brillouin zone scheme in the following.
The shaded areas in Figs.~\ref{fig5}(b), (c) and (d) are folded in the reduced Brillouin zone scheme.
Compared with the case of PM, additional discontinuities, whose magnitudes are $A_7(\bm{k})$, appear in $n_f^0(\bm{k})$ as shown in Fig.~\ref{fig5}(c). 
It is originated from the appearance of the AF order in $f$ electrons.
In the same way, additional discontinuities, whose magnitudes are $A_5(\bm{k})$, also appear in $n_c^0(\bm{k})$, 
although they are very small.

As shown in Fig.~\ref{fig5}(c), $n_f(\bm{k})$ is flattened in the whole Brillouin zone by the effect of $P^f_{\mathrm{G}}$ as in the case of PM.
However, discontinuities originated from $n_f^0(\bm{k})$ seems to be preserved in $n_f(\bm{k})$.
The discontinuities in $n_c(\bm{k})$ is rather uncertain because of their smallness (see the inset of Fig.~\ref{fig5}(d)). 
However, by fitting the data points by polynomials, we can find a finite jump at the original Fermi surface. 
In conclusion, we expect that the original Fermi surface is preserved through the projection also in AF$_\mathrm{h}$.

\begin{figure}[t!]
\begin{center}
\includegraphics[width=7.8cm]{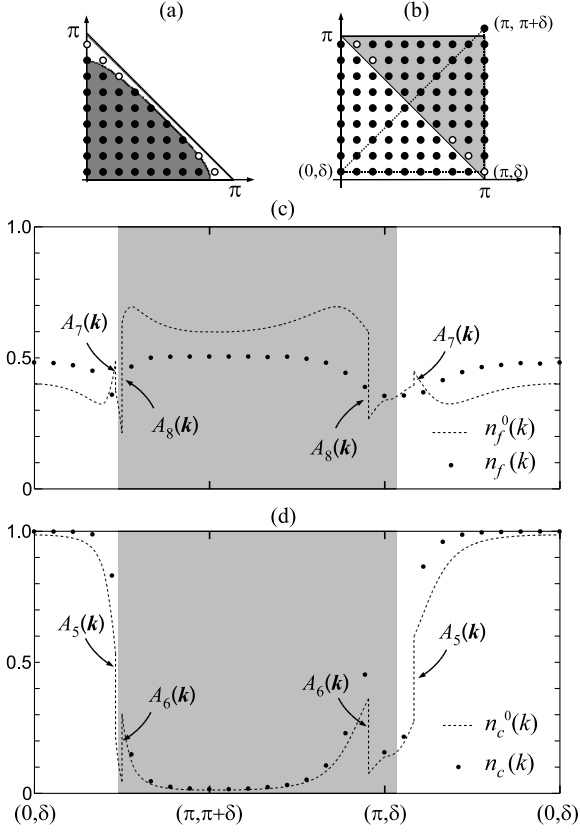}
\caption{Momentum distribution functions of AF$_{\mathrm{e}}$ at $E_F=-1.0$, $n=1.889$ and $V=0.90$. Notation is the same with that of Fig.~\ref{fig5}.
The optimized variational parameters are 
$(\tilde{V},\tilde{E_f},\Delta_f,\Delta_c)=(0.6607,-0.0905,0.4722,0.0000)$}
\label{fig6}
\end{center}
\end{figure}

Figure~\ref{fig6} shows the result for AF$_{\mathrm{e}}$ at $V=0.90$.
Contrary to the case of PM and AF$_\mathrm{h}$, the result is somewhat different in AF$_\mathrm{e}$. 
The discontinuities exist near the AF Brillouin zone boundary both in $n_c^0(\bm{k})$ and $n_f^0(\bm{k})$ as shown by the arrows in 
Figs.~\ref{fig6}(c) and (d). 
However, $n_f(\bm{k})$ is greatly flattened by projection and the discontinuities are greatly suppressed.
If these discontinuities disappear, $f$ electrons become completely localized and do not contribute to the Fermi surface. 
We cannot say whether the discontinuities in $n_f(\bm{k})$ really disappear or not within this calculation. 
However, it is certain that the itinerancy of $f$ electrons is greatly suppressed in AF$_\mathrm{e}$ compared with the case of PM and AF$_\mathrm{h}$.
On the other hand, the discontinuities in $n_c(\bm{k})$ seems to exist at the original Fermi surface, as in the case of PM and AF$_\mathrm{h}$.
In this sense, the volume and the shape of the Fermi surface is preserved through the projection also in AF$_\mathrm{e}$, whichever the discontinuities
in $n_f(\bm{k})$ disappear or not. 
This result shows that even if the $f$ electrons do not contribute to the Fermi surface, the $c$ electrons ``recognize'' the existence of $f$ electrons
and preserve the original Fermi surface.
It is possible that $c$ electrons are redistributed by the effect of projection and form a ``small'' Fermi surface whose volume is determined by the 
number of $c$ electrons alone. If it occurs, the discontinuities in $n_c(\bm{k})$ exist only in the unshaded area and the discontinuities in the
shaded area must disappear. But this is not the case.

\begin{figure}[t!]
\begin{center}
\includegraphics[width=7.2cm]{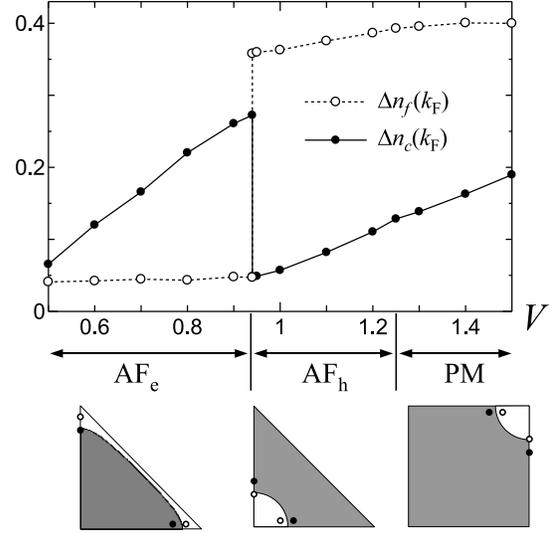}
\caption{$V$ dependences of the discontinuities of $n_c(\bm{k})$ and $n_f(\bm{k})$ at the Fermi surface. 
They are estimated from the average of two directions shown above.}
\label{fig7}
\end{center}
\end{figure}

Finally we discuss how much the electrons are correlated and become heavy from the discontinuities in $n_c(\bm{k})$ and $n_f(\bm{k})$.
We estimate the discontinuities at the Fermi surface from the difference between the values just inside and just outside of the 
Fermi surface.
As typical cases, we calculate them along two directions shown in the bottom panels in Fig.~\ref{fig7} and average their values.
The upper panel of Fig.~\ref{fig7} shows the obtained $\Delta n_c(k_{\mathrm{F}})$ and $\Delta n_f(k_{\mathrm{F}})$.
In PM and AF$_\mathrm{h}$, $\Delta n_c(k_{\mathrm{F}})$ and $\Delta n_f(k_{\mathrm{F}})$ continuously decrease as $V$ decreases. 
When the Fermi-surface reconstruction occurs (AF$_\mathrm{h}$ $\rightarrow$ AF$_\mathrm{e}$), a drastic change is observed.
$\Delta n_c(k_{\mathrm{F}})$ discontinuously increases since the position of $k_{\mathrm{F}}$ shifts near the AF Brillouin zone boundary, 
where the value of $n_c(\bm{k})$ is rather large compared with the former position. 
It shows that the $c$ electrons recover the weight at the Fermi energy.
Note that even after the Fermi-surface reconstruction, $c$ electrons are renormalized compared with the case of AF$_\mathrm{S}$
($\Delta n_c(k_{\mathrm{F}})=1$).
On the other hand, $\Delta n_f(k_{\mathrm{F}})$ discontinuously decreases to a small value through the Fermi-surface reconstruction. 
It shows that the $f$ electrons become almost localized and their itinerancy is fairly suppressed in AF$_\mathrm{e}$.
From these results, we expect that the transition to AF$_\mathrm{e}$ is a kind of orbital-selective Mott transition where only the $f$ electrons
become localized, as proposed in the CDMFT study~\cite{Leo}.
However, as mentioned before, we cannot say whether $\Delta n_f(k_{\mathrm{F}})$ becomes zero or not within this calculation. 
Moreover, the calculation for all directions is necessary for a precise discussion.
It is left for an interesting future problem.

\section{Discussion}
We discuss the relation between our results and the experiments.
If the Fermi-surface reconstruction occurs, changes in various physical quantities are expected.
The abrupt change of dHvA frequencies across the AF transition shows the appearance of a Fermi surface with different 
topology~\cite{Araki,Settai,Shishido1,Elgazzar},
indicating the occurrence of a Fermi-surface reconstruction. 
The picture that the $f$ electrons are localized in the AF phase and are itinerant in the paramagnetic phase, which is proposed from the comparison
of dHvA results and the band calculation, is also consistent with the scenario of a Fermi-surface reconstruction.

On the other hand, it have been proposed that the AF transition ($x\sim0.8$) and the abrupt change of dHvA frequencies ($x\sim0.4$) are not coincident 
in CeRh$_{1-x}$Co$_x$In$_5$. At $x=0.4$, the magnetic structure changes from incommensurate ($x<0.4$) to commensurate ($x>0.4$) within the AF 
phase~\cite{Goh}. 
The fitting with band calculation suggests that the former AF phase corresponds to CeRhIn$_5$ type (4$f$-localized picture) and the latter corresponds to
CeCoIn$_5$ type (4$f$-itinerant picture). We expect that this transition is closely related to the Fermi-surface reconstruction within the AF phase
(AF$_{\mathrm{e}}$ $\rightarrow$ AF$_{\mathrm{h}}$), although the more realistic calculation is necessary for a precise discussion.

The abrupt change of the Hall coefficient at the AF QCP in YbRh$_2$Si$_2$~\cite{Paschen} is also indicative of a Fermi-surface reconstruction.
However, the band calculation shows that the Hall coefficient is determined by several contributions with opposite signs and is very sensitive to the
energy level of $f$ electrons. It was pointed out that small changes in the $f$-electron occupation are sufficient to reproduce the experimental 
result~\cite{Norman}. 
Therefore, we cannot directly related the Fermi-surface reconstruction to the AF QCP in YbRh$_2$Si$_2$ at this stage.

We also refer to the effective mass in the AF state. The dHvA experiment shows that the Fermi surface of CeRhIn$_5$ is quite similar to 
that of non-4$f$ reference compound of LaRhIn$_5$~\cite{Shishido2}. 
It suggests that the contribution of the 4$f$ electrons to the Fermi surface is
rather small in CeRhIn$_5$. However, the obtained cyclotron mass in CeRhIn$_5$ is roughly by one order larger than that in LaRhIn$_5$~\cite{Shishido2}. 
We consider that the existence of 4$f$ electrons does induce this mass enhancement, keeping the Fermi surface almost unchanged. 
As we have shown in the case of AF$_{\mathrm{e}}$, it is possible that $f$ electrons do not contribute to the Fermi surface at all but the $c$ 
electrons are renormalized and heavy reflecting the existence of the $f$ electrons. It can explain the enhancement of cyclotron mass in CeRhIn$_5$.

\section{Summary}
In summary, we have studied the ground state phase diagrams and some physical quantities of the PAM in a two-dimensional square lattice with VMC
method. We have shown that
\begin{itemize}
\item{The Fermi-surface reconstruction which accompanies the localization of $f$ electrons occurs in a wide region of parameters. It induces the 
topological change of the Fermi surface and stabilizes the localized AF state.}
\item{The conventional itinerant AF transition also occurs separately from the Fermi-surface reconstruction.}
\item{Even in the localized AF state, the $c$ electrons are renormalized compared with the non-interacting case, reflecting the existence
of $f$ electrons.}
\end{itemize}
These results will be a good starting point to clarify the relation between the AF transition and the change of the Fermi surface in heavy-fermion
systems. For a more detailed discussion, calculations with realistic conditions will be necessary: three dimensional lattice structure, degeneracy of 
orbitals, magnetic ordering with incommensurate wave vector, and so on.
However, we think that the simple model discussed here capture the essence of the Fermi-surface reconstruction, which occurs in general as the 
evidence of strong localized character of $f$ electrons.

\section*{Acknowledgment}
The authors thank H. Tsunetsugu, K. Ueda, M. Imada, T. Sakakibara, M. Takigawa, S. K. Goh, M. Sutherland, E. Hassinger and 
F. F. Assaad for useful discussions. 
This work is supported by Grants-in-Aid from the Ministry of Education, Culture, Sports, Science and Technology of Japan (No. 16076203)
and also by a Next Generation Supercomputing Project, Nanoscience Program, MEXT, Japan.

\end{document}